\documentclass[reprint,aps,prd,showpacs,preprintnumbers,amsmath,amssymb]{revtex4-1}
\usepackage{graphicx}
\usepackage{graphics}
\usepackage{subfigure}
\usepackage{slashbox}
\usepackage{amsmath}
\usepackage{brackets}
\usepackage{psfrag}
\usepackage{epsfig}
\usepackage{lineno}
\usepackage{hyperref}
\usepackage{amssymb}
\begin{document}

\ProvideTextCommandDefault{\textonehalf}{${}^1\!/\!{}_2\ $}

\title{Comment on "Search for an axion-induced oscillating electric dipole moment for electrons using atomic magnetometers"}

\author{P.-H.~Chu}
\email[Email address: ]{pchu@lanl.gov}
\author{Y.~J.~Kim}
\email[Email address: ]{youngjin@lanl.gov}
\author{I.~Savukov}
\affiliation{Los Alamos National Laboratory, Los Alamos, New Mexico 87545, USA}
\date{\today}

\begin{abstract}
In the recent work~\cite{Chu:2018avg}, the authors proposed a new method measuring the electron oscillating electric dipole moment (eOEDM) using atomic magnetomaters. This eOEDM is induced by the interaction between the electron magnetic dipole moment, electric field and axion field. The result is sensitive to the axion-photon coupling according to Ref.~\cite{Hill:2015}. Here we want to describe that the same experimental method can be also sensitive to the axion-electron coupling according to Ref.~\cite{Alexander:2018}. In this article, we will show the corresponding sensitivity plot and compare with other constraints.
\end{abstract}
\pacs{32..Dk, 11.30.Er, 77.22.-d, 14.80.Va,75.85.+t}
\keywords{axion, dark matter}

\maketitle

In the recent paper~\cite{Chu:2018avg}, the authors have proposed a new method measuring the electron oscillating electric dipole (eOEDM) moment using atomic magnetometers. The eOEDM can be induced by the interaction between the electron magnetic dipole moment, electric field and axion field~\cite{Hill:2015, hill:2016, Hill:2017}. The result is sensitive to the axion-photon coupling $g_{a\gamma\gamma}$. The interaction can be derived by considering the magnetic dipole moment in the axion-modified Maxwell equations~\cite{Hill:2016zos}.

In Ref.~\cite{Alexander:2018}, the authors pointed out that the eOEDM can be also induced by the  axion-electron coupling $g_{aee}$. This requires the concern of the second order correction in the axion-modified Maxwell equations. The authors compared with the spin oscillation induced by the axion-wind as described in Ref.~\cite{graham:2013} and claimed that the eOEDM term becomes dominant if the axion mass is smaller than $10^{-6}$ eV giving the electric field around 1 keV/m. The amplitude of the eOEDM they derived is
\begin{align}
    d_e &= g_{aee}\frac{e}{m_e}\frac{\sqrt{2\rho_{DM}}}{m_a}\notag\\
    &= 8.3\times 10^{-23}~\text{e}\cdot\text{cm} (\frac{g_{aee}}{\text{GeV}^{-1}})(\frac{\text{eV}}{m_a})
\end{align}
where $g_{aee}$ is defined by $\lambda/m_e$ where $\lambda$ is the dimensionless Yukawa coupling~\cite{Alexander:2018}, $e$ is the electron charge, $m_e$ is the electron mass, $m_a$ is the axion mass and $\rho_{DM}\sim 0.3 \text{GeV}/\text{cm}^{3}$ is the dark matter density. Here we modify the Eq. 24 in Ref.~\cite{Alexander:2018} in order to make the coupling constant be consistent with Ref.~\cite{graham:2013} and other references.

\begin{figure}[b]
\includegraphics[width=0.5\textwidth]{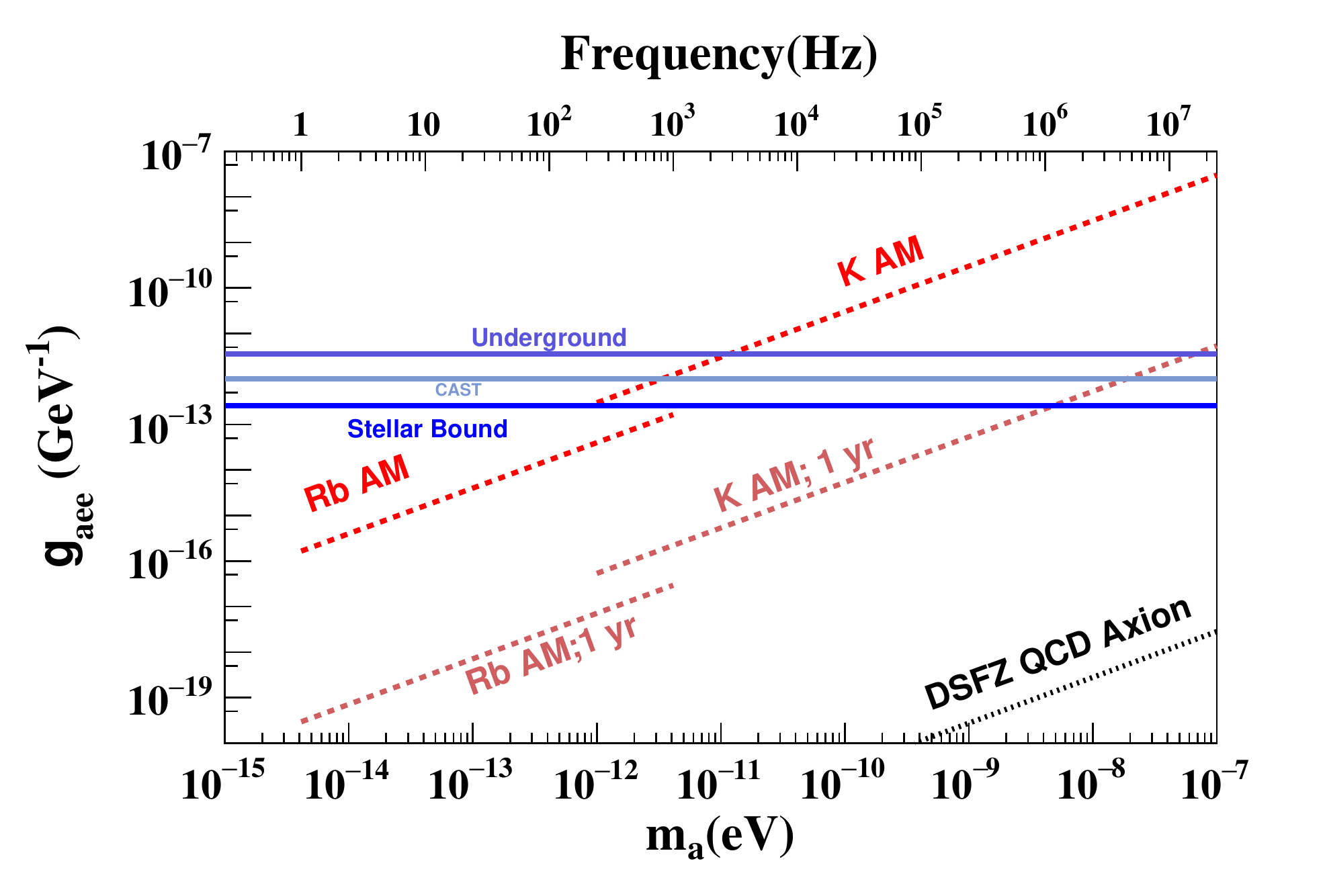}
\caption{Estimated sensitivity of our electron OEDM experiment based on Rb and K AMs to $g_{aee}$ on the axion mass range with 1 second integration time  and 1 year integration time (red dashed line). The constraints from the stellar bound, underground experiments and the DSFZ QCD axion are taken from Ref.~\cite{Irastorza:2018}. The constraint from CAST is from Ref.~\cite{Barth:2013}. }
\label{fig:gaee}
\end{figure}

The energy sensitivity of atomic magnetometers is about $2.9\times 10^{-20}$ eV~\cite{Chu:2018avg}. Therefore, the sensitivity of the energy shift due to the electron OEDM is limited by
\begin{align}
E R_{\text{K/Rb}} d_e=2.9\times 10^{-20}~\text{eV}
\end{align}
where $R_{\text{K/Rb}}$ is the EDM enhancement factor for K and Rb atoms, and $E=5$~kV/cm as described in Ref.~\cite{Chu:2018avg}. Figure~\ref{fig:gaee} shows the sensitivity for $g_{aee}$ using atomic magnetometers. The current constrains are all above $10^{-13}~\text{GeV}^{-1}$ while this method can improve the current sensitivity upto 3 orders when $m_a\sim 10^{-14}~\text{eV}$ for 1 second integration time and upto 6 orders for 1 year integration time.

In conclusion, the experimental setup in Ref.~\cite{Chu:2018avg} can be both sensitive to the axion-photon and axion-electron coupling. The recent white dwarf cooling measurement seemingly implied $g_{aee}\sim 1.6\times 10^{-13}$~\cite{Irastorza:2018,Giannotti:2017}. Additionally, some theories also prefer ultralight axion-like-particles (ALPs), which can have larger coupling than the QCD axion, as dark matter candidates~\cite{Arias:2012}. It is critical to improve the constraint of the axion coupling. 

\section*{ACKNOWLEDGMENTS}
The authors thank Dr. S. Alexander for useful discussion. This work was supported by the U.S. DOE through the LANL/LDRD program.

\bibliography{main}

\end{document}